\documentclass[a4paper]{article}    % LaTeX style for referee version
\usepackage{amsmath}
\usepackage{amssymb}
\usepackage{epsfig}
\usepackage{graphicx}
\usepackage{natbib}
\usepackage{subfigure}
\def\apj{ApJ}%
\def\apjl{ApJ}%
\def\aap{A\&A}%
\def\aaps{A\&AS}%
\def\solphys{Sol.~Phys.}%
\newcommand{\be}[1]{\begin{equation}\label{#1}}
\newcommand{\ee}{\end{equation}}
\begin{document}

%\titlerunning{Reversed granulation in the solar photosphere}

\author{M.~C.~M. Cheung\footnote{\emph{Present address:} Lockheed Martin Solar and Astrophysics
Laboratory, Bldg/252 3251 Hanover St., Palo Alto, CA 94304, USA.}, M. Sch\"ussler \\
Max Planck Institute for Solar System Research\\
37191 Katlenburg-Lindau, Germany\\
e-mail: [cheung,msch]@mps.mpg.de \\
\and F. Moreno-Insertis \\
Instit\'{u}to de Astrof\'{i}sica de Canarias\\
38200 La Laguna (Tenerife), Spain \\
Dept of Astrophysics, Faculty of Physics
\\University of La Laguna, 38200 La Laguna (Tenerife), Spain \\
e-mail: fmi@ll.iac.es}

\title{The origin of the reversed granulation in the solar photosphere}
\date{15 November 2006} \maketitle

\begin{abstract}We study the structure and reveal the physical nature of
the reversed granulation pattern in the solar photosphere by means
of 3-dimensional radiative hydrodynamics simulations. We used the
MURaM code to obtain a realistic model of the near-surface layers
of the convection zone and the photosphere. The pattern of
horizontal temperature fluctuations at the base of the photosphere
consists of relatively hot granular cells bounded by the cooler
intergranular downflow network. With increasing height in the
photosphere, the amplitude of the temperature fluctuations
diminishes. At a height of $z=130-140$ km in the photosphere, the
pattern of horizontal temperature fluctuations reverses so that
granular regions become relatively cool compared to the
intergranular network. Detailed analysis of the trajectories of
fluid elements through the photosphere reveal that the motion of
the fluid is non-adiabatic, owing to strong radiative cooling when
approaching the surface of optical depth unity followed by
reheating by the radiation field from below. The temperature
structure of the photosphere results from the competition between
expansion of rising fluid elements and radiative heating. The
former acts to lower the temperature of the fluid whereas the
latter acts to increase it towards the radiative equilibrium
temperature with a net entropy gain. After the fluid overturns and
descends towards the convection zone, radiative energy loss again
decreases the entropy of the fluid. Radiative heating and cooling
of fluid elements that penetrate into the photosphere and overturn
do not occur in equal amounts. The imbalance in the cumulative
heating and cooling of these fluid elements is responsible for the
reversal of temperature fluctuations with respect to height in the
photosphere.\end{abstract}

%\keywords{Convection - Radiative transfer - Sun: atmosphere - Sun:
%photosphere - Sun: granulation}

\section{Introduction}

In the quiet Sun, the most prominent photospheric feature is the
surface granulation pattern. Observations in white light show
bright granules with a typical size of $1$ Mm, embedded in a
network of dark intergranular downflow lanes and vertices. The
granulation pattern is non-stationary: studies of time-sequences
of the granulation pattern in the quiet Sun show that individual
features evolve with an average lifetime of about $5$
minutes~\citep[e.g.][]{Title:SolarGranulation}.

While intensity images in the visible continuum taken near
disk-center show bright granules and comparatively dark
intergranular boundaries, observations taken in the wings of
the~Ca~II~H \& K~lines exhibit a similar pattern, but with
reversed intensity
contrast~\citep*{Evans:Oddities,Suemoto:ReversedGranulation,Suemoto:ReversedGranulationII}.
This effect, known as the~\emph{reversed granulation}, has also
been detected in intensity measurements of photospheric spectral
lines~\citep{Espagnet:PenetrationOfSolarGranulation,Balthasar:IntensityCorrelationsInGranularSpectra,
Kucera:FeILinesInQuietSolarPhotosphere,RodriguezHidalgo:ReversedGranulation,JanssenCauzzi:ReversedGranulation}.
These observations suggest that the reversal of the intensity
contrast occurs already at a height of about $150$ km above the
mean geometric level where $\tau_{\rm 500} = 1$.

Recently,~\citet*{Rutten:ReversedGranulationInCaIIH} acquired
simultaneous, speckle-reconstructed images of quiet-Sun
granulation in the G-band and the~Ca~II~H~line. On the basis of
cross-correlations between time sequences of intensity images in
these two spectral ranges they found that the anticorrelation
between the two patterns is highest if the~Ca~II~H~intensity maps
are time-delayed by $2-3$ minutes with respect to the G-band
intensity maps. By computing anticorrelation coefficients between
(1) intensity images in the continuum and in the ~Fe~I~7090~
\AA~line and (2) intensity images in the continuum and in the wing
of
the~Ca~II~8542~\AA~line,~\citet{JanssenCauzzi:ReversedGranulation}
also found that the strongest anticorrelation occurs for a time
delay of $\sim 2$ min.~\citet*{Leenaarts:ReversedGranulation}
extended the work of~\citet*{Rutten:ReversedGranulationInCaIIH} by
calculating synthetic intensity images in the blue continuum
and~in~the~Ca~II~H~wing (in the LTE approximation) from a
3-dimensional radiative hydrodynamic simulation. From the
unsmoothed synthetic intensity images, they found that the
anticorrelation between the continuum images and the~Ca~II~H~wing
images does not vary significantly between a time delay of $0$ and
$2$ minutes. The anticorrelation decreased for longer time delays.
When the images were smoothed to $1.5$ arcsec resolution, a
time-delay of $2-3$ minutes yielded a significantly higher value
of the anticorrelation compared to zero time delay.
~\citet*{Rutten:ReversedGranulationInCaIIH} speculated that
magnetic fields play no major role in the formation of the
reversed granulation. The results
of~\citet{Leenaarts:ReversedGranulation} is evidence supporting
this view.

Numerical simulations have firmly established the solar
granulation as a radiative-convective
phenomenon~\citep{Nordlund:ConvectionAndWaveGeneration,Nordlund:SolarConvection,Steffen:SolarGranulation,SteinNordlund:SolarGranulation}.
Granules are upwellings of hot, high-entropy material originating
from the convection zone and overshooting into the stably
stratified photosphere. Radiative losses make the overshooting
material relatively cold and dense. Its overturning motion
supplies the network of intergranular downflows. Simulations have
also revealed that the pattern of horizontal temperature
fluctuations in the middle/upper photosphere is inverted compared
to the corresponding pattern at the base of the photosphere.
Whereas at the base of the photosphere, the granules are
relatively hot compared to the intergranular network, the former
are relatively cooler for $\tau_{\rm 500}\lesssim 0.3$. It is well
known that this reversal in the temperature fluctuations in the
photosphere is responsible for the reversal in the observed
intensity
pattern~\citep*{Nordlund:ConvectionAndWaveGeneration,Suemoto:ReversedGranulation,Leenaarts:ReversedGranulation}.

What is less clear is the physical mechanism which leads to the
reversal of the temperature fluctuations with height. It is
sometimes incorrectly assumed that fluid travelling in the
optically thin layers of the photosphere does so adiabatically. As
pointed out by~\citet{Nordlund:ConvectionAndWaveGeneration}, this
cannot be true because material with an original temperature of
$5900$ K penetrating and expanding adiabatically in the
photosphere would reach a temperature of $1800$ K after traversing
only three pressure scale heights. Such low temperatures are not
detected in the
photosphere.~\citet{SteinNordlund:SolarGranulation}~indeed find
that fluid rising in the photosphere is typically reheated by the
radiation field. This leads one to conclude that non-adiabatic
effects, such as radiative energy exchange, must play a crucial
role in the formation of the reversed temperature fluctuations.

The aim of the present paper is to investigate the physical origin
of the reversal of temperature fluctuations. As we shall see in
the following sections, radiative energy exchange is indeed key to
the temperature fluctuation reversal. By following in detail the
time histories of fluid elements penetrating into the photosphere,
we identify how radiative heating and cooling, in tandem with the
horizontal pressure fluctuations, causes the reversed granulation.
The paper is structured as follows.~In~\S~\ref{sec:setup}, we
describe the set of governing equations, the numerical methods
used and the simulation setup. In~\S~\ref{sec:results}, we discuss
results from the numerical simulation, including the mean
stratification of the 3D model
(\S~\ref{subsec:mean_stratification}), the temperature structure
in the photosphere
(\S~\ref{subsec:reversed_granulation_structure}) and reveal the
physical origin of the reversed granulation
(\S~\ref{subsec:origin_reversed_granulation}). Finally, in
\S~\ref{sec:conclusion}, we discuss how the present work dispels
previous misconceptions and improves our understanding of the
phenomenon.

\section{Simulation setup}
\label{sec:setup}
\subsection{Governing equations}\label{subsec:equations}
For a realistic treatment of the dynamics and energetics in the
near-surface layers of the non-magnetic solar atmosphere, it is
important to consider the effects of compressibility, energy
exchange via radiative transfer and partial
ionization~\citep{SteinNordlund:SolarGranulation,Rast:IonizationI,Rast:SolarIonization,Voegler:PhD}.
The hydrodynamics equations, the radiative transfer equation (RTE)
and the equation of state (EOS) including the effects of partial
ionization constitute the system of governing equations for our
model. Since we are confining our attention to the near-surface
layers of the convection zone and the photosphere, the
approximation of local thermodynamic equilibrium (LTE) suffices
for the treatment of the radiative transfer. The hydrodynamics
equations include the continuity equation, the momentum equation
and the energy equation, viz.
\begin{eqnarray}
\frac{\partial \varrho}{\partial t} & + &  \nabla \cdot (\varrho
\mathbf{v})=0\label{eqn:continuity},\\
\frac{\partial \varrho \mathbf{v}}{\partial t} & + & \nabla \cdot
\left( \varrho \mathbf{v}\otimes\mathbf{v}\right ) = -\nabla p +
\varrho \mathbf{g} + \nabla
\cdot \underline{\underline{\tau}},\label{eqn:momentum}\\
\frac{\partial \varrho \varepsilon}{\partial t} &+& \nabla\cdot
\left (
\varrho\mathbf{v}\left[\varepsilon+\frac{p}{\varrho}+\frac{1}{2}|\mathbf{v}|^2
 \right] \right )= \nonumber\\
& & \nabla\cdot(\mathbf{v}\cdot\underline{\underline{\tau}}) +
\nabla\cdot(K\nabla T) + \varrho (\mathbf{g}\cdot\mathbf{v}) +
Q_{\rm rad} \label{eqn:energy},
\end{eqnarray}
\noindent where $\mathbf{v}$ is the fluid velocity, $\varrho$ the
mass density and $\varepsilon$ the internal energy density per
unit mass. The operator $\otimes$ denotes the tensor product. In
the near-surface layers of the Sun, the gravitational acceleration
is almost constant and we take $\mathbf{g} = -2.74\times10^4
\hat{z}$ cm s$^{-2}$.

The hydrodynamics
equations~(\ref{eqn:continuity})-(\ref{eqn:energy}) must be
supplemented by the appropriate constitutive relations describing
the material properties of the gas. These include the equation of
state (EOS), functions specifying the viscous stress tensor
$\underline{\underline{\tau}}$ and thermal conductivity $K$, as
well as functions specifying the radiative properties of the gas
(e.g. source function $S_\nu$, opacity $\kappa_\nu$ etc.). The
frequency-integrated radiative heating term appearing on the
r.h.s. of Eq. (\ref{eqn:energy}) is given by
\begin{equation}
Q_{\rm rad} = \int 4\pi\kappa_\nu\varrho(J_\nu - \bar{S}_\nu){\rm
d}\nu, \label{eqn:qrad_two}
\end{equation}
\noindent where $J_\nu$ is the angle-averaged specific intensity
and $\bar{S}_\nu$ the angular-averaged source function.

\subsection{Numerical method}

For our radiative hydrodynamics simulations, we have used
the~\emph{M}PS/\emph{U}niversity of Chicago \emph{Ra}diative
\emph{M}HD (MURaM) code~\citep*{Voegler:MURaM}. In the absence of
magnetic fields, the equations solved by MURaM reduce to the
equations given in the preceding section.

In the following, we briefly describe the numerical methods
implemented in MURaM. For further details, we refer the reader
to~\citet{Voegler:PhD}~and~\citet{Voegler:MURaM}. MURaM integrates
the discretized equations
(\ref{eqn:continuity})-(\ref{eqn:energy}) on a 3-dimensional
cartesian grid, with constant grid-spacing in each cartesian
direction. The spatial discretization of the equations is based on
a fourth-order centered-difference scheme on a five-point stencil.
Time-stepping is carried out using an explicit fourth-order
Runge-Kutta scheme, with the maximum size of the time step
determined by a modified CFL-criterion.

In order to prevent the nonlinear cascade of energy towards the
grid scale, artificial diffusivities are implemented to stabilize
the simulations. A combination of~\emph{hyper-}
and~\emph{shock-resolving} diffusivities are
used~\citep{Caunt:Hyperdiffusivities}. Hyperdiffusivities are
implemented in such a way, that the diffusion terms associated
with them are comparable in magnitude to the inertial terms only
at length scales close to the grid-spacing. For the thermal and
viscous coefficients $K$ and $\nu$, the hyperdiffusivities are
defined using the scheme described in~\citet{Voegler:MURaM}.

The RTE is solved with the short-characteristics method on a 3D
cartesian grid, with the grid centers of this radiative grid
defined by the cell corners of the grid used for discretizing the
hydrodynamics equations (\ref{eqn:continuity})-(\ref{eqn:energy}).
For each point on this radiative grid, the RTE is solved over $24$
ray directions ($3$ per octant). The first moment of $I_\nu$ over
these rays provides the radiative flux density $\mathbf{F}_\nu$,
which in turn allows $Q_{\rm rad}$ to be calculated.

MURaM can treat the radiative transfer in the grey or non-grey
case~\citep[see][and references therein]{Voegler:Nongrey}. For
this study, we have carried out two simulations, namely a grey run
and a non-grey run. The results presented in the rest of this
paper correspond to the non-grey case. The main difference between
the results of the two cases it that the non-grey case yields
smaller temperature fluctuations in the photosphere than the grey
case. This difference arises because the grey case underestimates
the smoothing effects of radiative transfer in the optically thin
layers of the photosphere in comparison with the non-grey
case~\citep*{Voegler:Nongrey}. Our explanation for the reversed
granulation phenomenon (see
\S~\ref{subsec:reversed_granulation_structure}), however, applies
to both cases.

The EOS relations $T=T(\varepsilon,\varrho)$ and
$p=p(\varepsilon,\varrho)$ are used by MURaM in the form of
look-up tables. The look-up tables were evaluated by solving the
Saha-Boltzmann equations. The first ionization of the $11$ most
abundant elements was considered. For a detailed account of how
the tables were evaluated, we refer the reader to Appendix A
of~\citet{Voegler:MURaM}.

\subsection{Simulation setup}
In order to obtain a thermally relaxed 3D model of the upper
convection zone and the photosphere, we have chosen the following
setup. The horizontal size of the simulation domain is $24$ Mm
$\times12$ Mm, with each direction spanned by $480$ and $240$ grid
cells, respectively. The height of the domain is $2.3$ Mm, and is
spanned by $144$ grid cells. The grid-spacing in the horizontal
and vertical directions is $50$ and $16$ km, respectively.
Periodic boundary conditions are assumed for the vertical side
boundaries. The bottom boundary is open and allows mass to flow
smoothly into and out of the domain~\citep[see][]{Voegler:MURaM}.

For the initial condition, we took the 1D height profiles of
$\varrho$ and $\epsilon$ from the mixing length model
of~\citet*{Spruit:ConvectionZone} and introduced a plane-parallel
stratification into the simulation domain. To break the symmetry,
we imposed pressure perturbations on the order of a few percent.
We then ran the simulation until we obtained a thermally relaxed,
statistically stationary state.

\section{Simulation results}
\label{sec:results}
\subsection{Mean stratification}
\label{subsec:mean_stratification} 
\begin{figure}
\centering
\includegraphics[width=0.7\textwidth]{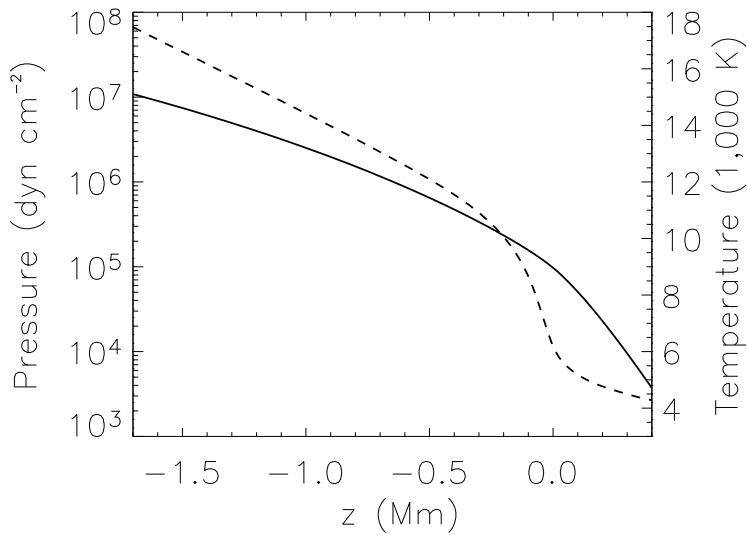}
\caption{Horizontally averaged pressure (solid) and temperature
(dashed) as functions of height $z$.}\label{fig:pt_profiles}
\end{figure}

The horizontally averaged
profiles of pressure and temperature as functions of height are
shown in Fig.~\ref{fig:pt_profiles}, respectively, as solid and
dashed curves. The geometrical height $z=0$ corresponds to the
geometrical level where the continuum optical depth at $500$ nm
is, on average, unity. The height dependence of the mean profiles
are in general agreement with the simulation results
of~\citet{SteinNordlund:SolarGranulation}.

\begin{figure}
\centering
\includegraphics[width=0.7\textwidth]{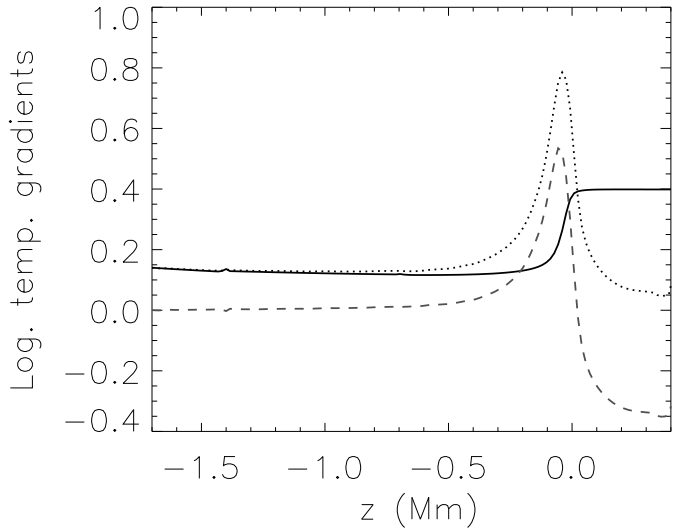}
\caption{Logarithmic temperature gradients
 $\nabla$ (dotted curve), $\nabla_{\rm ad}$ (solid black)
    and the superadiabaticity $\delta=\nabla-\nabla_{\rm ad}$ (dashed curve).}
    \label{fig:nabla_profiles}
\end{figure}

Figure~\ref{fig:nabla_profiles} shows the logarithmic temperature
gradients determined from the mean profiles:
\begin{eqnarray}
\nabla &:=& \frac{{\rm d}\ln{T}}{{\rm d}\ln{p}}, {\rm~and} \\
\nabla_{\rm ad} &:=&
\left(\frac{\partial\ln{T}}{\partial\ln{p}}\right)_s.
\end{eqnarray}
\noindent$\nabla$ (dotted curve) is the actual average temperature
gradient in the simulation and $\nabla_{\rm ad}$ (solid black
curve) is the adiabatic temperature gradient. $\nabla_{\rm ad}$
describes the variation of temperature of a fluid element
undergoing adiabatic expansion or compression.

The stability of the stratification to convective motion is
affected by changes in the ionization state. The superadiabaticity
of the stratification is defined as $\delta := \nabla -
\nabla_{\rm ad}$ and is plotted as a dashed curve in
Fig.~\ref{fig:nabla_profiles}. In the photosphere, the chemical
species are almost completely monatomic. For such a gas,
$\gamma=5/3$ and hence, $\nabla_{\rm ad} = 1-1/\gamma=0.4$. With
increasing depth into the convection zone, the ionization fraction
of hydrogen (and traces of other species) increases. This has the
effect of decreasing the adiabatic temperature gradient, so that
$\nabla_{\rm ad }< 0.4$ while $\delta$ remains positive. In other
words, changes in the ionization state of the gas reduces
$\nabla_{\rm ad}$ and thus enhances buoyancy driving.

\subsection{The reversal of temperature fluctuations in the middle photosphere}
\label{subsec:reversed_granulation_structure}
\begin{figure*}
\centering
\includegraphics[width=0.9\textwidth]{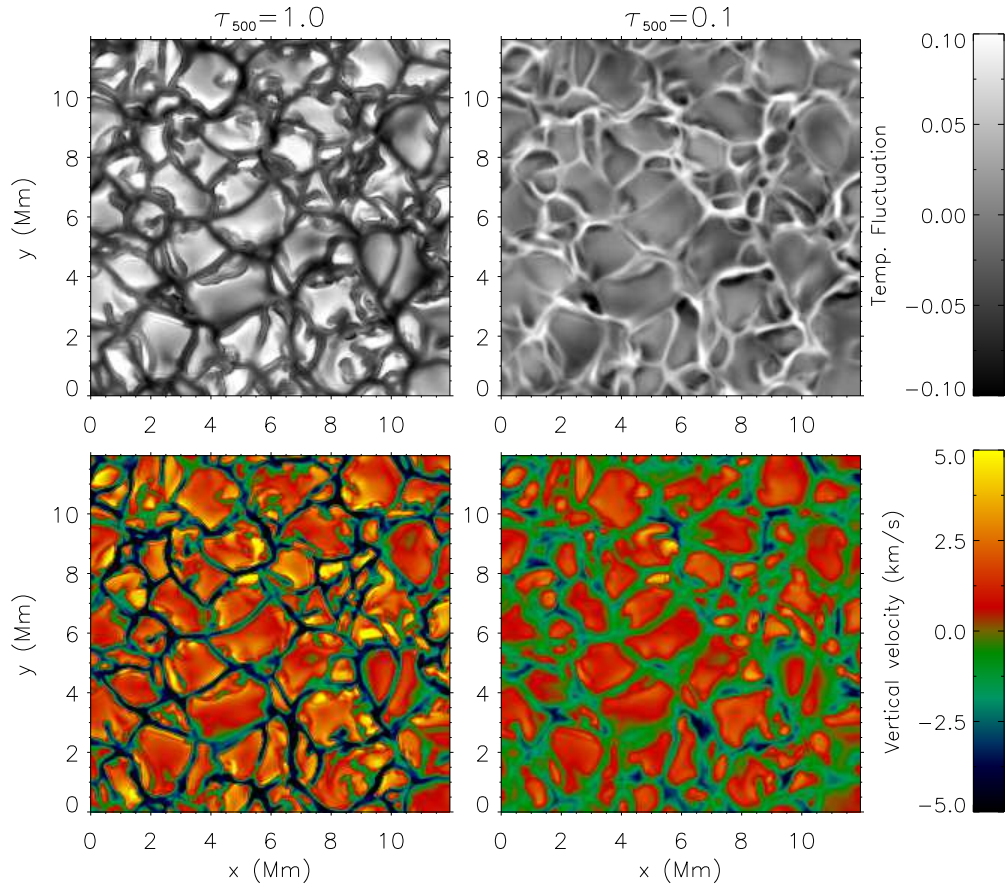}
\caption{Temperature fluctuations ($\Delta T/\bar{T}$, upper row)
and vertical velocity (lower row) at two surfaces of constant
optical depth. The left column shows the patterns at $\tau_{\rm
500}=1$, the right $\tau_{\rm 500}=0.1$. The reversed granulation
corresponds to a reversal of the temperature fluctuations while
the velocity pattern remains qualitatively
unchanged.}\label{fig:reverse_granulation}
\end{figure*}
In this section, we study the reversed granulation in our 3D
model. For lines that form under approximate LTE conditions, such
as some Fe I lines and the wings of the~Ca~II~H~\&~K~lines,
fluctuations in line intensity are a proxy for diagnosing
temperature fluctuations. The intensity image therefore maps the
temperature distribution in the layer of the atmosphere in which
the corresponding part of the line is formed. An inspection of the
temperature fluctuations at different heights of the atmosphere
allows one to determine the height at which the temperature
fluctuations start to reverse.

The upper row of Fig.~\ref{fig:reverse_granulation} shows the
relative temperature fluctuations ($\Delta T/\bar{T}$) at the
surfaces $\tau_{\rm 500}=1$ (left column) and $\tau_{\rm 500}=0.1$
(right column). The lower row gives the vertical velocity at the
same surfaces. On average, the $\tau_{\rm 500}=0.1$ surface is
elevated above the $\tau_{\rm 500}=1$ surface by $160$ km. At
$\tau_{\rm 500}=1$, one finds the normal granulation pattern: the
granules are hotter than the intergranular lanes. At $\tau_{\rm
500}=0.1$, the reversed granulation is already clearly
discernible: the lanes are regions of temperature excess. When we
inspect the velocity structure, we find that the pattern at
$\tau_{\rm 500}=0.1$ is very similar to that at $\tau_{\rm
500}=1$, albeit with smaller flow speeds. The upflows in the
higher layers result from the overshooting of upwelling gas from
the convection zone. Thus, the reversal of the pattern of
temperature (and intensity) fluctuations is~\emph{not} accompanied
by a reversal in the velocity pattern~\citep[e.g.
see][]{RuizCobo:AverageSolarGranule}.

\begin{figure}
\centering
\includegraphics[width=0.7\textwidth]{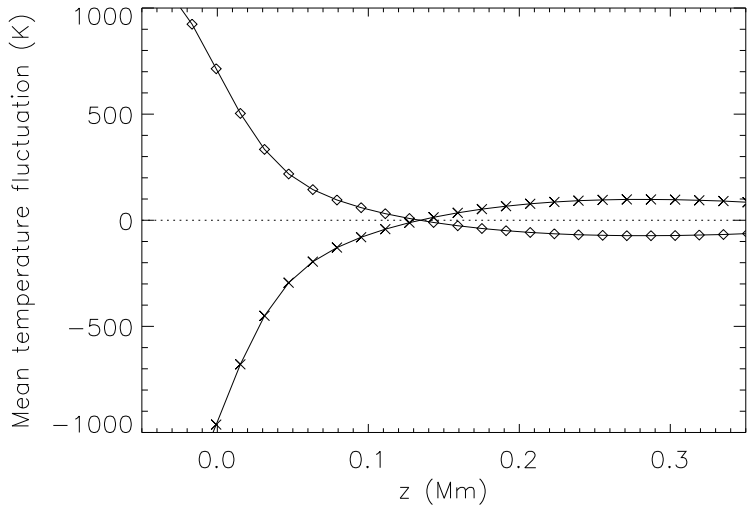}
\caption{Mean temperature fluctuation ($\Delta T$) of granules
(diamonds) and of intergranular lanes (crosses) as a functions of
geometrical height. At $z=130-140$ km, the cells and boundaries
have, on average, the same temperature. Above this geometrical
height, the intergranular lanes are relatively hotter than the
granules.}\label{fig:reversal_geometric}
\end{figure}

In order to determine the geometrical height at which the reversed
granulation begins, we consider the vertical velocity and
temperature as functions of $z$ for vertical lines of sight
(l.o.s.). Distinguishing the l.o.s. profiles according to the sign
of the vertical velocity at $\tau_{\rm 500}=1$, we separately
determine averages for profiles of the temperature fluctuations in
granules and in the intergranular downflow network, respectively.
The fluctuations are evaluated relative to the overall average
temperature profile. These two quantities are plotted as functions
of $z$ in Fig.~\ref{fig:reversal_geometric}. Diamonds indicate the
mean temperature fluctuation in the granules whereas crosses
indicate the mean temperature fluctuation in the intergranular
lanes. These two are not simply the negative of each other because
the cells have a larger area fraction than the boundaries. In the
convection zone and in the lower photosphere, upflowing material
in the granules are relatively hot compared to downflowing
material in the lanes. The reversal of the horizontal temperature
fluctuations occurs, on average, in the middle photosphere at a
height of $z=130-140$ km. Above that height, the intergranular
lanes are hotter than the granules with a temperature excess of
$100$-$200$ K.

\begin{figure}
\centering
\includegraphics[width=0.7\textwidth]{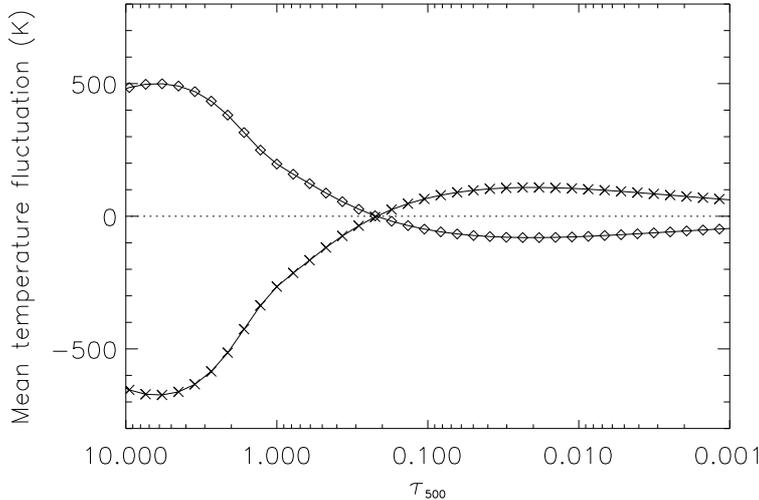}
\caption{Mean temperature fluctuation ($\Delta T$) of granules
(diamonds) and of intergranular lanes (crosses) as a functions of
optical depth. The reversed granulation appears at optical depths
of about $\tau_{\rm
500}=0.2$.}\label{fig:reverse_temperature_fluctuation}
\end{figure}

We have also determined the temperature profiles in granules and
in the intergranular network as functions of optical depth along
vertical lines of sight. The average temperature fluctuations of
the granules and the intergranular lanes as functions of
$\tau_{\rm 500}$ are indicated by diamonds and crosses
respectively in Fig.~\ref{fig:reverse_temperature_fluctuation}. In
the convection zone ($z<0$), the temperature fluctuations at a
given geometrical height are much larger than the temperature
fluctuations at constant optical depth. As pointed out
by~\citet{SteinNordlund:SolarGranulation}, this is due to the high
temperature sensitivity of the H$^-$ contribution to the continuum
opacity. In this figure, the two curves cross each other at
$\tau_{\rm 500} = 0.22$. On average, the surface of constant
optical depth at $\tau_{\rm 500} = 0.22$ is elevated above the
$\tau_{\rm 500}=1$ surface by $120$ km.

The fact that the two different averaging procedures yield
somewhat different heights ($z=130-140$ km and $z=120$ km,
respectively) for the layer of the photosphere where the
temperature fluctuations reverse their sign is easily understood.
In the former case, we calculated the horizontal temperature
fluctuations relative to the mean temperature of the layer at
different geometrical heights. In the latter case, the temperature
fluctuations correspond to deviations from the mean temperature at
surfaces of constant optical depth. Since these surfaces are
corrugated (e.g. the rms fluctuation of the geometrical height of
the $\tau_{\rm 500}=1$ surface is $30$ km), temperature
fluctuations across such a surface only approximate the horizontal
temperature fluctuations at the same geometrical height. In terms
of both geometrical height and optical depth, our determination of
the level at which the reversal occurs is in good agreement with
previous observational
work~\citep{RuizCobo:AverageSolarGranule,RodriguezHidalgo:ReversedGranulation,Borrero:TwoComponentModel,Puschmann:HiReseGranulationII}.

\subsection{The origin of the reversal of temperature fluctuations}
\label{subsec:origin_reversed_granulation}

The analysis in the previous section confirmed the reversal of
horizontal temperature fluctuations in the middle photosphere. Now
we proceed to explain the physics behind this phenomenon. There
are two important aspects to the explanation. Firstly, we need to
understand the nature of convective overshoot of granular fluid
into the photosphere: namely, what is the balance of forces
controlling the dynamics of convective penetration and overturn?
Secondly, we need to understand the thermodynamical processes
experienced by the overshooting material.

In order to address these issues, we have examined the histories
of tracer fluid elements in our simulation. At time $t=0$
(corresponding to the snapshot shown in
Fig.~\ref{fig:reverse_granulation}), tracer fluid elements were
introduced just beneath the surface inside a granule and
propagated according to the simulated velocity field. The physical
properties of the fluid elements (e.g. $\mathbf{v}, \varrho,
\varepsilon, s, Q_{\rm rad}$ etc.) at their actual positions were
determined by tricubic interpolation from the values in the
adjacent cells~\citep{Lekien:Tricubic}.

\begin{figure*}
\includegraphics[width=\textwidth]{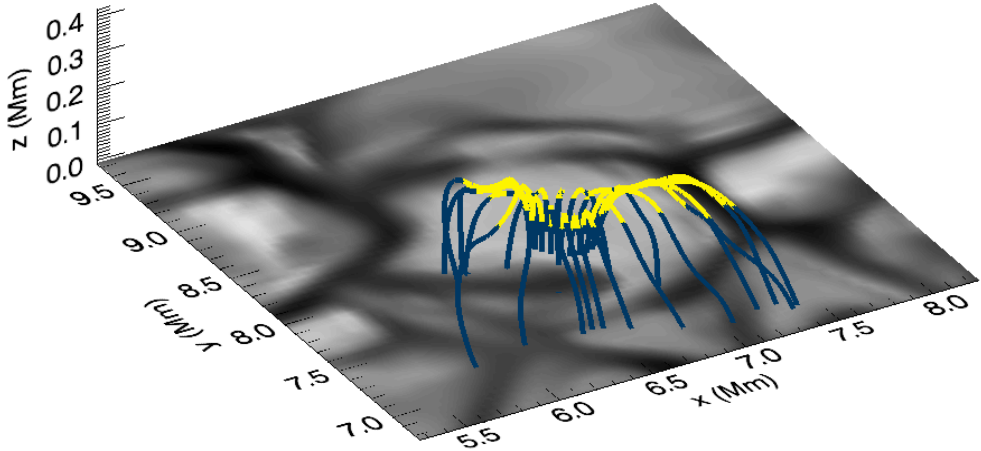}
\caption{Trajectories of tracer fluid elements in the photospheric
neighbourhood of a granule. The greyscale image in the $z=0$ plane
shows the vertical velocity there at the time when the tracer
fluid elements were released. The colour coding of the
trajectories corresponds to the sign of the radiative heating
$Q_{\rm rad}$ experienced by the fluid elements (dark blue means
$Q_{\rm rad}<0$, yellow means $Q_{\rm
rad}>0$).}\label{fig:fountain_Qtot}
\end{figure*}

Figure~\ref{fig:fountain_Qtot} shows a subvolume of the simulation
domain in the neighbourhood of the granule centered at
$[x,y]=[7,8]$ Mm in Fig.~\ref{fig:reverse_granulation}. The
greyscale image at $z=0$ indicates the vertical velocity at $z=0$
at the instant when the tracer fluid elements were released
($t=0$). The colour-coding of the trajectories indicate the sign
of $Q_{\rm rad}$: dark blue means $Q_{\rm rad}<0$ (cooling),
yellow means $Q_{\rm rad}>0$ (heating). The set of trajectories
resembles a fountain: the fluid elements originate from the
granule interior, travel up to the photosphere and eventually
overturn and descend into the downflow network in the convection
zone.

\subsubsection{The motion of fluid elements in the photosphere}

First we concentrate on the motion and dynamics of the fluid
elements. As Fig.~\ref{fig:fountain_Qtot} shows, in spite of the
strongly stable average stratification, the fluid elements are
able to penetrate into the photosphere. The shape of the
trajectories tell us that in the photosphere, they experience a
deceleration in the vertical direction and a deflection towards
the edge of the granule.

In the following, we give a brief description of the force balance
which drives the horizontal granular outflows and the feeding of
the downflows in the intergranular lanes. Inspection of the
horizontal pressure fluctuations in the photospheric layers of our
simulation reveal that local pressure hills in the interior of
granules are responsible for the buoyancy braking of the upflows
and the deflection of material horizontally outwards from the
centres of granules. At the interface between two granules, the
horizontal outflows from the granules meet and develop localized
pressure ridges. These localized regions of pressure enhancement
serve to retard the horizontal granular outflows. So, when a fluid
element in the granular outflow approaches the intergranular
boundary, it is decelerated by the pressure gradient. Finally, the
pressure enhancement at the boundaries, together with gravity,
work in tandem to accelerate the fluid element downwards, thereby
feeding the intergranular downflow network. These results are in
accordance with results from 2D idealized simulations of
compressible
convection~\citep*{Hurlburt:Magnetoconvection,Hurlburt:PenetrativeConvection}
and 3D near-surface convection
simulations~\citep{Nordlund:SolarConvection,SteinNordlund:SolarGranulation}.

\subsubsection{The thermodynamic history of fluid elements in the photosphere}
\label{subsec:thermodynamic_history} To avoid confusion in the
following discussion, at this point we would like to clarify the
following terminology: By heating (cooling), we refer to a
non-adiabatic process which increases (decreases) the specific
entropy of a fluid element. For example, when a fluid element
absorbs more radiation than it emits, it is heated and gains
entropy. A fluid element which is compressed adiabatically,
however, will have a higher final temperature but is not heated.
In the absence of viscous dissipation or thermal conduction, the
equation governing the evolution of the specific entropy of a
fluid element is
\begin{equation}
 \frac{Ds}{Dt} = \frac{Q_{\rm rad}}{\varrho T},\label{eqn:entropy_evolution}
\end{equation}
\noindent where $D/Dt$ denotes the Lagrangian
derivative~\citep{MihalasAndMihalas:Foundations}. The
corresponding equation governing the temperature of a fluid
element is
\begin{eqnarray}
\frac{D\ln{T}}{Dt} &=& (\gamma_3-1) \frac{D\ln{\varrho}}{Dt} +
\left(\frac{\partial \ln{T}}{\partial s}\right)_\varrho
\frac{Ds}{Dt}\label{eqn:temperature_evolution} \\
& = &-(\gamma_3-1) \nabla \cdot \mathbf{v} + \left(\frac{\partial
\ln{T}}{\partial s}\right)_\varrho \frac{Q_{\rm rad}}{\varrho
T}.\label{eqn:temperature_evolution2}
\end{eqnarray}
\noindent where $\gamma_3 = 1+\left( \frac{\partial
\ln{T}}{\partial \ln{\varrho}}\right)_s >1$ is Chandrasekhar's
third adiabatic exponent. Two effects contribute to changes in the
gas temperature of a fluid element. The first term on the r.h.s.
of Eq. (\ref{eqn:temperature_evolution2}) describes the role of
adiabatic expansion or compression. Clearly, the effect of
adiabatic expansion ($\nabla \cdot \mathbf{v} > 0,~Q_{\rm rad}=0$)
is to decrease the gas temperature. The second term on the r.h.s.
of Eq. (\ref{eqn:temperature_evolution2}) describes the role of
radiative heating. As expected, radiative heating (cooling) leads
to an increase (decrease) of the gas temperature. The relative
importance of the two effects in controlling the fluid temperature
depends on their characteristic timescales. Inspection of Eq.
(\ref{eqn:temperature_evolution2}) motivates us to define the
dynamical and radiative timescales respectively as
\begin{eqnarray}
t_{\rm dyn} &=& \left | \frac{1}{(\gamma_3-1)\nabla \cdot
\mathbf{v}} \right|,{\rm~and} \\
t_{\rm rad} &=& \left| \left(\frac{\partial s}{\partial
\ln{T}}\right)_\varrho \frac{\varrho T}{Q_{\rm rad}}\right|.
\end{eqnarray}

\noindent In the case that $t_{\rm dyn} \ll t_{\rm rad}$, the
motion of a fluid element is essentially adiabatic. In the
opposite extreme, where $t_{\rm dyn}\gg t_{\rm rad}$, the fluid
element is always in radiative equilibrium with its surroundings.

Now we proceed to discuss the thermodynamic history of fluid
elements overshooting into the photosphere. Owing to the large
opacity for $z<0$, fluid in the upwellings in the convection zone
ascend almost adiabatically. The decrease of a fluid element's
temperature during its rise in the convection zone is due to
adiabatic expansion. As it approaches the photosphere, the opacity
drops rapidly and the fluid element loses entropy by radiative
cooling. Thereafter, the fluid element overshoots into the stably
stratified photosphere. Although the fluid element is now
optically thin, it is not completely transparent to radiation and
continues to exchange energy with its surroundings. In regions
above granules, (consisting of overshooting material), radiative
energy exchange predominantly leads to heating. This effect can be
seen in the colour-coding of the trajectories in
Fig.~\ref{fig:fountain_Qtot}. This figure clearly indicates that
as the fluid elements penetrate into the photosphere (and as they
are diverted toward the intergranular lanes), they experience
radiative heating.

\citet{SteinNordlund:SolarGranulation} also found that rising
fluid in the photosphere is systematically reheated by the
radiation field (see Fig. 18 of their paper). The reason for this
behaviour is the thermostatic action of the radiation field to
keep the fluid elements close to radiative equilibrium. The
radiative equilibrium temperature $T_{\rm RE}$ of a fluid element
embedded in a radiation field is defined as the temperature at
which the fluid element emits and absorbs radiation in equal
amounts (i.e. $Q_{\rm rad}= 0$). One can make a simple estimate
for the~\emph{lower} limit of $T_{\rm RE}$ by assuming the fluid
element to be located at $\tau=0$. In the approximation of grey
radiative transfer, $\kappa_\nu$ is replaced by the mean opacity
$\bar{\kappa}$, so that Eq. (\ref{eqn:qrad_two}) becomes
\begin{equation}
Q_{\rm rad } = 4\pi \bar{\kappa} \varrho
[J-B(T)],\label{eqn:qrad_mean_kappa}
\end{equation}
\noindent where $B(T) = \int_0^\infty B_\nu (T){\rm d}\nu =
(\sigma/\pi) T^4$ is the frequency-integrated Planck function and
$J=(4\pi)^{-1}\int_{4\pi} \int_0^\infty I_\nu {\rm d}\nu {\rm
d}\Omega$ is the angle-averaged, frequency-integrated intensity.
Assume that the solar atmosphere has plane-parallel symmetry. In
this case, $I_\nu$ is symmetric about the $z$-axis. Making use of
the Eddington-Barbier approximation, we obtain
\begin{eqnarray}
\nonumber J &=& \frac{1}{4\pi}\int_0^\infty \oint_{4\pi} I_\nu
{\rm d}\Omega {\rm d}\nu\\
\nonumber  & = & \int_0^\infty \left[ \frac{1}{2} \int_{-1}^{1}
B_\nu(T|_{\tau=\mu}) {\rm d}\mu \right ] {\rm d}\nu
\\
& = & \frac{\sigma}{2\pi} \int_{-1}^{1} (T|_{\tau=\mu})^4 {\rm
d}\mu,\label{eqn:J_expression}
\end{eqnarray}
\noindent where $\mu=\vec{\mu}\cdot\mathbf{\hat{z}}=\cos{\theta}$.
Here $\mathbf{\mu}$ is the direction cosine vector and $\theta$ is
the angle between $\mathbf{\mu}$ and $\hat{z}$. By definition,
$T_{\rm RE}$ is such that $B(T_{\rm RE}) = (\sigma/\pi) T_{\rm
RE}^4 = J$. To evaluate $J$ from Eq. (\ref{eqn:J_expression}), we
use the average temperature profile $\bar{T}(\tau)$ from our
model. This yields $T_{\rm RE}=4600$ K. This result tells us why
granular material ascending in the photosphere~\emph{must} be
radiatively heated. Consider a fluid element that is rising in the
photosphere. If the fluid element rises adiabatically, its
temperature would drop so quickly with height, that it becomes
cooler than the radiative equilibrium temperature. As soon as this
happens, the fluid element will absorb more radiation than it
emits ($Q_{\rm rad} > 0$). By the same token, consider a fluid
element in the photosphere which has been transported from the
interior of a granule towards an intergranular lane. As it
descends, it is compressed. If it were to descend adiabatically,
its temperature would soon be above $T_{\rm RE}$, which means it
must then cool radiatively.
\begin{figure*}
\centering
\includegraphics[width=0.85\textwidth]{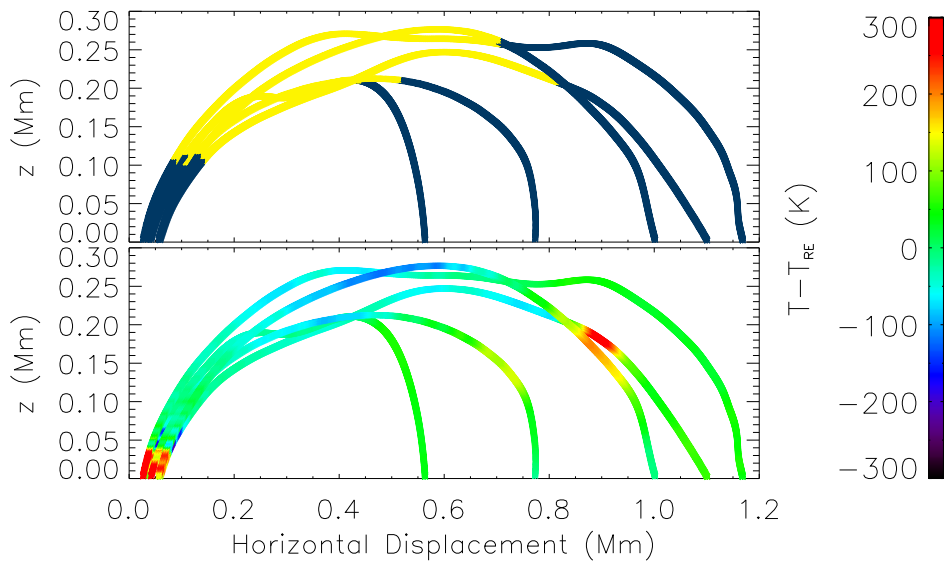}
\caption{Upper panel: Trajectories are colour-coded according to
the sign of $Q_{\rm rad}$ (dark blue and yellow correspond to
$Q_{\rm rad}<0$ and $Q_{\rm rad}>0$, respectively). Lower panel:
Trajectories are colour-coded according to the value of $T-T_{\rm
RE}$, the deviation of the gas temperature from the radiative
equilibrium temperature. In both panels, the fluid elements
emanate from the left-hand-side and travel towards the
right.}\label{fig:deviation_from_t_re}
\end{figure*}

Figure~\ref{fig:deviation_from_t_re} provides an illustration of
how overshooting fluid elements become cooler than $T_{\rm RE}$ as
they penetrate into the photosphere, and subsequently become
hotter than $T_{\rm RE}$ as they descend towards the convection
zone. In the figure, the height $z$ of a fluid element is plotted
as a function of the horizontal displacement from its original
position (i.e. $[\Delta x^2+\Delta y^2]^{1/2}$). In both panels,
the fluid elements travel from the left to the right of the plot.
In the upper panel, the trajectories are colour-coded according to
the sign of $Q_{\rm rad}$. In the lower panel, they are
colour-coded according to the instantaneous, local value of
$T-T_{\rm RE}$. $T_{\rm RE}$ is evaluated by keeping track of
$J_i$ and $\kappa_i$ ($i$ refers to the index of the frequency
bin) along the trajectories and then finding a temperature $T$
such that $\sum \kappa_i B_i(T) = \sum \kappa_i J_i$.

As a rising fluid element approaches the base of the photosphere,
its temperature can be up to $400$ K higher than $T_{\rm RE}$.
This causes the fluid element to radiate intensely. Once the fluid
element has reached the optically thin layers of the photosphere,
adiabatic expansion and radiative heating compete to control the
temperature of the fluid element. The relative importance of these
two effects (measured by the ratio of $t_{\rm dyn}$ and $t_{\rm
rad}$) determine whether the fluid element will be driven towards
or away from radiative equilibrium. For a fluid element
penetrating into the middle photosphere in a granular environment,
we find that $0.1\lesssim t_{\rm dyn}/t_{\rm rad} \lesssim 1$.
This means that the decrease in fluid temperature due to expansion
of the fluid element is only partially compensated by radiative
heating. Consequently, its temperature decreases and it departs
further away from $T_{\rm RE}$. In
Fig.~\ref{fig:deviation_from_t_re}, we can see a general trend for
$T-T_{\rm RE}$ to become more negative as a fluid element rises.
For fluid elements that penetrate beyond $z= 250$ km, $T-T_{\rm
RE}$ can reach a value of $-200$ K.

\begin{figure}
\centering
\includegraphics[width=0.7\textwidth]{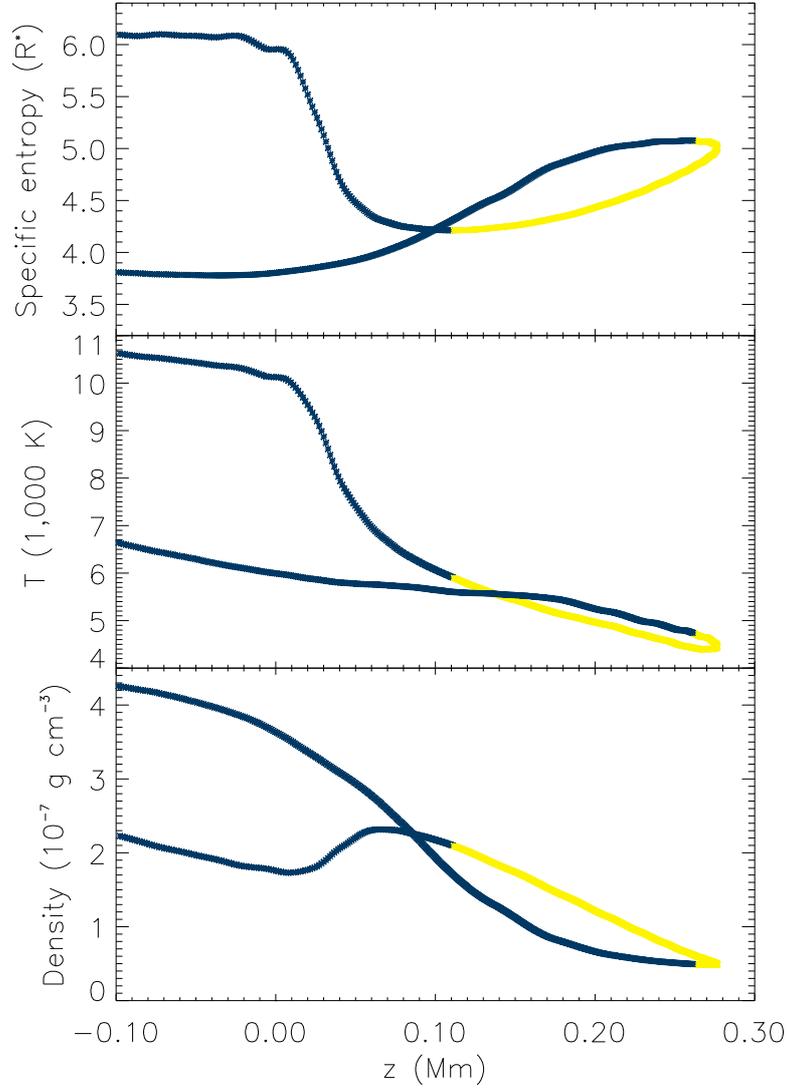}
\caption{Profiles of specific entropy $s$, temperature $T$ and
density $\varrho$ along one of the fluid element trajectories
shown in Fig.~\ref{fig:fountain_Qtot}. As the fluid element
initially emerges at the base of the photosphere ($z\approx0$), it
has relatively high entropy, high temperature and low density. The
colour-coding indicates the sign of $Q_{\rm rad}$ experienced by
the fluid element along its trajectory (dark blue means $Q_{\rm
rad}<0$, yellow means $Q_{\rm rad}>0$). Above $z=140$ km, the
fluid element is hotter during descent than during its ascent.
This is a result of radiative heating of the fluid element in the
photosphere.} \label{fig:hysteresis_curves}
\end{figure}

Figure~\ref{fig:hysteresis_curves} shows the profiles of $s$, $T$
and $\varrho$ along one of the trajectories shown in
Fig.~\ref{fig:fountain_Qtot}. The colour-coding indicates the sign
of $Q_{\rm rad}$ experienced by the fluid element along its
trajectory (dark blue means $Q_{\rm rad}<0$, yellow means $Q_{\rm
rad}>0$). The motion of the fluid element before it emerged at the
surface is essentially adiabatic, with $s=6.1 R^\star$, where
$R^\star$ is the universal gas constant. As it approaches
$z\approx 0$, radiative cooling leads to a steep drop of $s$ and
$T$ with height. Accompanying this cooling is a simultaneous
compression of the fluid element, so that it becomes temporarily
denser than the fluid beneath it. Thereafter, the fluid element
continues to ascend and the density decreases again. When it has
reached $z=0.1$ Mm, the fluid element is cooler than $T_{\rm RE}$
and it begins to be heated radiatively. This radiative heating
leads to a rise in entropy, while the temperature continues to
decrease because the effect of expansion overwhelms that of
radiative heating. Finally, at $z=0.28$ Mm, the fluid element
overturns and begins its descent. At this moment, the fluid
element has a gas temperature below $T_{\rm RE}$, so that it
continues to be heated radiatively even though its temperature
rises by compression. Only at $z=0.26$ Mm ($\approx 20$ km below
the apex of the fluid element's trajectory) is the temperature of
the fluid element above $T_{\rm RE}$ and $Q_{\rm rad}$ becomes
negative.

The profiles of $s$, $T$ and $\varrho$ in
Fig.~\ref{fig:hysteresis_curves} exhibit hysteresis. There are two
reasons for this behaviour. Firstly, there is a lag between when
the fluid element overturns and when $Q_{\rm rad}$ switches sign
(see previous paragraph). Secondly, the asymmetry in the speed of
upflows (slower) and downflows (faster) means that the cumulative
heating experienced by the fluid element over its journey upwards
in the photosphere is greater than the accumulative radiative
cooling it undergoes on the return journey. These two factors
contribute to the temperature and entropy enhancements of the
intergranular boundaries (downflows) with respect to the granules
(upflows).
\begin{figure}
\centering
\includegraphics[width=0.7\textwidth]{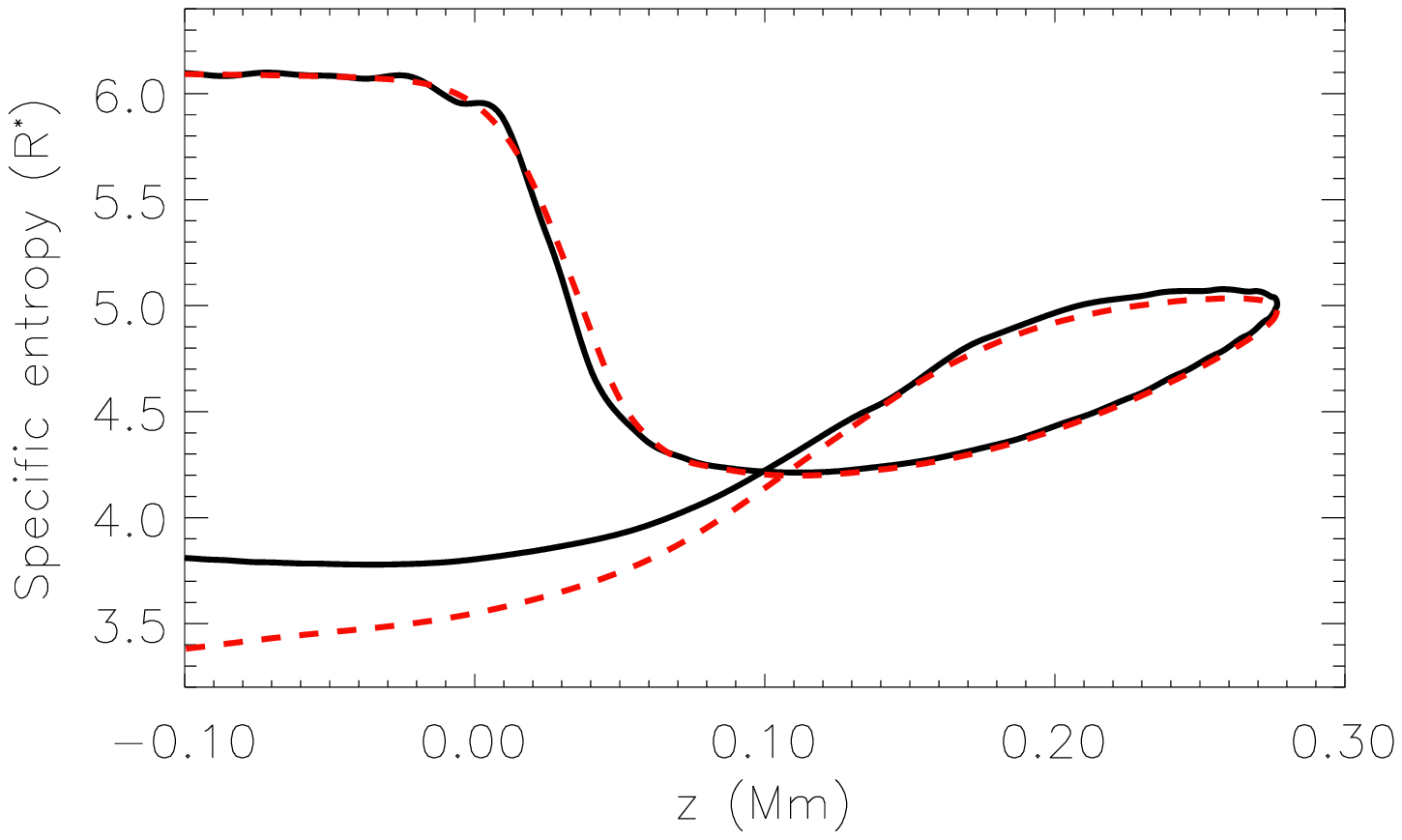}
\caption{Reconstruction of specific entropy profile. The black
solid curve shows the entropy profile as obtained by following the
fluid element in the simulation (same as in the upper panel of
Fig.~\ref{fig:hysteresis_curves}). The red dashed curve shows the
reconstructed entropy profile, $s_{\rm recon}$, calculated by
integrating Eq. (\ref{eqn:entropy_evolution}). }
\label{fig:hysteresis_prediction}
\end{figure}

In order to test whether dissipative effects (viscous and thermal,
e.g. in shocks) play a role in causing the temperature enhancement
in the intergranular lanes in the upper photosphere, we have
reconstructed the profiles of $s$ along the trajectory neglecting
hyperdiffusivity effects. By keeping track of $Q_{\rm rad}(t)$,
$\varrho(t)$ and $T(t)$ along the trajectory of a fluid element,
we can reconstruct the time history of $s$ for a fluid element.
Given these profiles and the initial entropy $s_0$ of the fluid
element, Eq. (\ref{eqn:entropy_evolution}) can be integrated to
yield the reconstructed entropy profile $s_{\rm recon}(t)$. This
quantity is plotted in Fig.~\ref{fig:hysteresis_prediction}
together with the entropy profile shown in
Fig.~\ref{fig:hysteresis_curves}. The good match of the two
profiles in the layers $z\gtrsim 0$ tells us that radiation is
indeed the cause of hysteresis in the entropy and temperature
profiles. The reconstructed profile $s_{\rm recon}$ begins to
deviate from the tracer profile once the fluid element approaches
the narrow turbulent downflow jets in the convection zone. Here
the radiative effects diminish and dissipative effects described
by the hyperdiffusive terms become important.

\section{Conclusion}
\label{sec:conclusion} The reversed granulation is a reversal of
the pattern of intensity contrast between intensity maps in the
continuum and intensity maps in spectral lines forming in the
middle to upper photosphere. It has long been established that
this reversal of the intensity fluctuations is a result of the
sign change of the horizontal temperature fluctuations (see
Fig.~\ref{fig:reverse_temperature_fluctuation}). Analysis of our
3D numerical model of the near-surface convection zone and
photosphere confirms the existence of this reversal in the
temperature fluctuations above a height of $z=130-140$ km, in
agreement with previous observational findings.

By following in detail the time histories of fluid elements
overshooting into the photosphere, we were able to identify the
processes that determine the horizontal temperature fluctuations
in the photosphere. We find that changes in the temperature of
fluid elements in the photosphere result from adiabatic expansion
and radiative heating. Whereas the former decreases the gas
temperature of a fluid element, the latter increases it towards
the radiative equilibrium temperature $T_{\rm RE}$. The ratio of
the associated characteristic timescales tells us that, for a
fluid element rising in the optically thin layers, the decrease of
temperature due to expansion is only partially cancelled by
radiative heating. This disparity allows a fluid element to depart
from radiative equilibrium as it penetrates into the photosphere.
Due to this deviation from $T_{\rm RE}$, a fluid element
overturning in the optically thin layers still experiences some
radiative heating as it begins to descend. This, together with the
fact that downflows have higher vertical speed than upflows, leads
to a hysteretic behaviour of the thermodynamic history of the
fluid element (see Fig.~\ref{fig:hysteresis_curves}).

In various guises in previous works, the following explanation has
often been invoked to explain the origin of the reversal of
temperature fluctuations: since the photosphere is
subadiabatically stratified, material rising from granules into
this stable layer has lower temperature that its `mean
surroundings'. This view is problematic in two respects. Firstly,
it implicitly assumes that material penetrating into the
photosphere does so in an adiabatic fashion. Secondly, it is not
at all clear what are the `mean surroundings' of a fluid element
embedded in convective flow consisting of disjointed upwellings
separated by an intergranular network. Our present efforts
highlight the importance of radiative energy exchange in the
dynamics of the photosphere and helps dispel possible
misconceptions regarding the origin of the reversed granulation in
the photosphere.
\\ \\
\noindent \textbf{Acknowledgements} \\
\noindent MCMC thanks Prof. Franz Kneer and his group at the
Georg-August University of G\"{o}ttingen for valuable discussions
on this topic. MCMC acknowledges financial support from the
International Max Planck Research School at the Max Planck
Institute for Solar System Research.

%\bibliography{6390manu}
%\bibliographystyle{aa}

\end{document}